\documentclass[12pt]{article}
\usepackage{graphicx}
\begin{document}
\pagestyle{myheadings}

\newcommand{\be}{\begin{equation}}
\newcommand{\ee}{\end{equation}}
\bibliographystyle{unsrt}
\vspace{0.5cm}

\begin{center} {\Large A Thought Construction of Working Perpetuum Mobile
of the Second Kind}\\ \vspace{0.5cm}

V. \v{C}\'{a}pek and J. Bok\\ \vspace{0.5cm}

Institute of Physics of Charles University, Faculty of Mathematics
and Physics,\\ Ke Karlovu 5, 121 16 Prague 2, Czech
republic\\(Tel. (00-420-2)2191-1330 or (00-420-2)2191-1450,
\\Fax (00-420-2)296-764,
\\E-mail capek@karlov.mff.cuni.cz or bok@karlov.mff.cuni.cz)
\vspace{0.5cm}

March 17, 1999

\end{center}
\vspace{0.5cm}

A previously published model of the isothermal Maxwell demon as one of
models of open quantum systems endowed with faculty of selforganization is
reconstructed here. It describes an open quantum system interacting with a
single thermodynamic bath but otherwise not aided from outside. Its activity
is given by the standard linear Liouville equation for the system and bath.
Owing to its selforganization property, the model then yields cyclic
conversion of heat from the bath into mechanical work without compensation.
Hence, it provides an explicit thought construction of perpetuum mobile
of the second kind, contradicting thus the Thomson formulation of the
second law of thermodynamics. No approximation is involved as a special
scaling procedure is used which makes the kinetic equations employed
exact.
\vspace{1cm}


Submitted to Czech. J. Phys.; preliminary version reported as a poster at\\
MECO 23, April 27-29, 1998, ICTP, Trieste
\newpage

\markright{V. Capek and J. Bok: Construction of Perpetuum Mobile of the Second Kind}

{\bf 1. Introduction}

In this letter, we should like to report on a result which can be
obtained, for the model in question, without approximations from
standard quantum theory of open systems governed by the linear Liouville -
von Neumann equation. Irrespective of this, it contradicts the second law
of thermodynamics. Leaving details of physical motivation to a next
publication, we mention here just the fact that the main inspiration for
construction of the model has been taken from biology, namely from
topological changes of biologically important molecules upon detecting, at
a specific site (receptor), particles (excitations, molecules or molecular
groups) to be processed \cite{Good}.

Previous version of the model has been published in \cite{Cap1}. Detailed
treatment of this model \cite{Cap2} as well as other microscopic models of
open quantum systems working on analogous principles
\cite{Cap4,Cap5,CapBok,CapTri,CapMan} revealed property of spontaneous
(i.e. not induced by external flows) selforganization.  This then leads to
such unexpected phenomena contradicting basic principles
of statistical thermodynamics as, e.g., violation of consequences of the
detailed balance (in connection with impossibility of rigorous
justification thereof). From this, implicit violations of the second
law of thermodynamics could be deduced as mentioned in, e.g.,
\cite{Cap5,CapBok}. Here, we reconstruct the model so that it is able to work
cyclically and without compensation as a perpetuum mobile of the second
kind, in the sense contradicting explicitly the Thomson formulation of the
second law of thermodynamics \cite{Kub}.

{\bf 2. Model}

The fully quantum Hamiltonian of our model in its simplest version
(see also \cite{Cap7}) can be as usual written as a sum of the
Hamiltonians of the (extended) system, thermodynamic bath, and that of the
system-bath interaction. Thus,
\be H=H_S+H_B+H_{S-B} \label{totham} \ee
where
\[ H_S=J(c_{-1}^{\dagger}c_0+c_0^{\dagger}c_{-1})\cdot
|d\rangle\langle d|+I(c_1^{\dagger}c_0+c_0^{\dagger}c_1)\cdot
|u\rangle\langle u|+K(c_1^{\dagger}c_{-1}+c_{-1}^{\dagger}c_1)\]
\be +\frac{\epsilon}{2}[1- 2c_0^{\dagger}c_0]\cdot[|u\rangle\langle
u|-|d\rangle\langle d|] \label{HamS}. \ee
As for the Hamiltonians of the bath $H_B$ and that of the system-bath
interaction $H_{S-B}$, they are not important here. Before explaining the
symbols used in (\ref{HamS}), let us only add that we shall assume
$H_{S-B}$ to consist of two additive and non-interfering contributions
$H_{S-B}'$ and $H_{S-B}''$. Here $H_{S-B}'$ causes transitions between
states of the central system $|u\rangle$ and $|d\rangle$ (see below) while
$H_{S-B}''$ is responsible for a sufficiently intense (and, for the sake
of simplicity, equally strong) dephasing of the particle states at all three
sites. As an example, one can take a model of the bath consisting of
harmonic oscillators $H_B=\sum_k\hbar\omega_kb_k^{\dagger}b_k$
interacting with the central system by a linear
site-local coupling causing the $|u\rangle\leftrightarrow|d\rangle$
transitions exactly as in \cite{Cap1} but complemented, for our purposes, by,
e.g., a non-interfering site-local coupling of the particle to the
bath with the same coupling constant at all three sites. This means, e.g.,
$H_{S-B}=\frac{1}{\sqrt{N}}\sum_k\hbar\omega_kG_k(b_k
+b_{-k}^{\dagger})[|u\rangle\langle d|+|d\rangle\langle u|]$ $+\frac{1}{N}
\sum_{kk'}\hbar\sqrt{\omega_k\omega_{k'}}g_{kk'}(b_k+b_{-k}^{\dagger})
(b_{k'}+b_{-k'}^{\dagger})
(c_{-1}^{\dagger}c_{-1}+c_0^{\dagger}c_0+c_{+1}^{\dagger}c_{+1})$ $\equiv
H_{S-B}'+H_{S-B}''$.
These specific forms of $H_B$ and $H_{S-B}$ will not be, however, used below.

As for $H_S$ in (\ref{HamS}), we have chosen the simplest model with only three
particle states, i.e. sites ($m=-1,0$ and $+1$). Operators $c_m^{\dagger}$
and $c_m$, $m=-1,0,$ or $+1$ designate the particle creation and annihilation
operators at site $m$. For simplicity, we shall assume only one particle in the
system. This is why we do not need the commutational (anticommutational)
relations of the particle creation and annihilation operators. In (\ref{HamS}),
$I$, $J$ and $K$ are transfer (hopping or resonance) integrals connecting the
sites involved. Worth noticing is that in (\ref{HamS}), the forth and back
transfers in any pair of the sites are always with the same amplitude as a
consequence of the hermicity of $H_S$. The one-directional character of the
process reported here is not owing to a contingent difference between these
amplitudes but results exclusively (as it will become clear later on) from
the existence of spontaneous processes. Site `0' is understood to be attached
to a central system representing, e.g., a specific molecule or molecular
group (e.g. tail connecting site `0' with either site `-1' or `+1' but not
both simultaneously). This
central system is assumed to have (in a given range of energies of
interest) two levels with energies $\pm\epsilon/2$ (with corresponding
states $|d\rangle$
and $|u\rangle$; we assume $\epsilon>0$). At the moment when the particle is
transferred to site `0' attached to the central system, the relative order
(on the energy axis) of the
two levels of the central system gets interchanged. (This of course causes
instability of the central system with respect to the $|d\rangle\rightarrow
|u\rangle$ transition which is the same effect as additional-load-induced
instability of a ship in water.) Asymmetry in transfer rates
$|u\rangle\leftrightarrow|d\rangle$ is ensured by the above spontaneous
processes with respect to the bath. Conversion of this up-and-down
asymmetry into the left-and-right one is then owing to the above special
form of the first two terms on the right hand side of (\ref{HamS}) discussed
below. Technically, this imbalancing is due to the fourth term on the right
hand side of (\ref{HamS}) proportional to $1-2c_0^{\dagger}c_0$ and may be
in reality due to correlation effects as the particle transferred may, upon
its transition to site $0$, change the topology
(originally stable conformation may become energetically disadvantageous) or
orientation of the central system in space.
(Such changes condition activity of many biologically important molecules
in living organisms \cite{Good}.)

{\bf 3. Equations of motion}

In our theory, we shall closely follow \cite{Cap2} but shall be interested
in just a stationary situation. That is why we can apply any proper kinetic
method yielding time derivatives of matrix elements of the density matrix
of our extended system (particle + the central two-level system), and to
set these time derivatives zero. In order to avoid unnecessary technical
complications, we have avoided time-convolution methods. From mutually
equivalent time-convolutionless methods, we have chosen (like in \cite{Cap2})
that of Tokuyama and Mori \cite{ToMo}. In order to avoid unnecessary
discussions about role of approximations, we have employed one of the
scaling procedures turning, in the limiting sense, Tokuyama-Mori equations
into exact kinetic equations. This means the
following steps: \begin{itemize} \item We formally introduce a joint small
parameter, say $g$, of both $H_{S-B}$ and all the hopping integrals $I$,
$J$, and $K$ in the sense of setting $H_{S-B}\propto g$ but $I,\,J,\,
K\propto g^2$. Right here, let us remind the reader of the fact that this
scaling of some parameters from $H_S$ is
what distinguishes our approach from scaling standardly used in weak
coupling theories (inapplicable to our presumably intermediate or rather
strong coupling situation discussed here) \cite{Davi}. \item We introduce
new time unit $\tau\propto g^2$, i.e. introduce new time $t'=t/\tau_0$.
This step formally disappears as far as we are, like here, interested in
just the stationary (long-time) asymptotics. \item We divide our general
Tokuyama-Mori kinetic equations by $g^2$ and perform the limit
$g\rightarrow 0$. \end{itemize}
The reader can easily see that this method preserves just the lowest order
(in $g^2$) terms which can be calculated exactly. The result can be
reported as follows (for details see a next extended publication):

First, we arrange all the 36 matrix elements $\rho_{\alpha\gamma}(t)$ of
the density matrix of our extended system ($\alpha,\,\beta...=md$ or $mu$
with $m=0$ or $\pm1$ while $d$ or $u$ designate the states of the central
system) in groups of nine designating
\[(\rho_{uu})^T=\biggl(\rho_{-1u,-1u},\;\rho_{0u,0u},\;\rho_{1u,1u},\;
\rho_{-1u,0u},\;\rho_{-1u,1u},\;\rho_{0u,-1u},\;\rho_{0u,1u},\;
\rho_{1u,-1u},\; \rho_{1u,0u}\biggr),\]
\[(\rho_{dd})^T=\biggl(\rho_{-1d,-1d},\;\rho_{0d,0d},\;\rho_{1d,1d},\;
\rho_{-1d,0d},\;\rho_{-1d,1d},\;\rho_{0d,-1d},\;\rho_{0d,1d},\;
\rho_{1d,-1d},\; \rho_{1d,0d}\biggr),\]
\be (\rho_{ud})^T=\biggl(\rho_{-1u,-1d},\;\rho_{0u,0d},\;
\rho_{1u,1d},\;\rho_{-1u,0d},\;\rho_{-1u,1d},\;\rho_{0u,-1d},\;
\rho_{0u,1d},\;\rho_{1u,-1d},\; \rho_{1u,0d}\biggr) \label{rhonn} \ee
and similarly for $\rho_{du}$. Superscript $\cdots^T$ designates
transposition. Then the above kinetic equations in the asymptotic time
domain (and after the above scaling) read as
\[\left(\begin{array}{c} 0 \\ 0 \\ 0 \\
0 \end{array}\right)=\left(\begin{array}{cccc}{
\cal A} & {\cal B} & {\bf 0} & {\bf 0} \\ {\cal C} & {\cal D} & {\bf
0} & {\bf 0} \\ {\bf 0} & {\bf 0} & \cdots & \cdots \\
{\bf 0} & {\bf 0} & \cdots & \cdots \end{array}\right)\cdot
\left(\begin{array}{c}\rho_{uu} \\ \rho_{dd} \\ \rho_{ud} \\
\rho_{du} \end{array}\right).\] \be \label{Equat2} \ee
Here, in the square matrix, all the elements are in fact blocks
$9\times9$. Hence, the whole set splits into two independent sets of
$18$ equations each; we shall be interested just in that one for
$\rho_{uu}$ and $\rho_{dd}$. This reads as in (\ref{Equat2}) with typical
forms of the block $9\times9$ matrices ${\cal A,B,C}$ and ${\cal D}$.
Here
\[{\cal A}=\] \[\left(\begin{array}{ccccccccc}
-\Gamma_{\downarrow}& 0 & 0 & 0 & iK/\hbar & 0 & 0 & -iK/\hbar & 0 \\
0 & -\Gamma_{\uparrow} & 0 & 0 & 0 & 0 & iI/\hbar & 0 & -iI/\hbar \\
0 & 0 & -\Gamma_{\downarrow} & 0 &-iK/\hbar& 0 & -iI/\hbar
& iK/\hbar & iI/\hbar\\
0 & 0 & 0 & k-2\Gamma & iI/\hbar & 0 & 0 & 0 & -iK/\hbar \\
iK/\hbar& 0 &-iK/\hbar& iI/\hbar & -\Gamma_{\downarrow}-2\Gamma& 0 & 0 & 0 & 0 \\
0 & 0 & 0 & 0 & 0 & k^*-2\Gamma &iK/\hbar& -iI/\hbar & 0 \\
0 & iI/\hbar & -iI/\hbar & 0 & 0 &iK/\hbar& k^*-2\Gamma & 0 & 0 \\
-iK/\hbar& 0 & iK/\hbar & 0 & 0 & -iI/\hbar & 0 & -\Gamma_{\downarrow}-2\Gamma & 0
\\ 0 & -iI/\hbar & iI/\hbar &-iK/\hbar& 0 & 0 & 0 & 0 & k-2\Gamma
\end{array}\right). \]
\be \label{BlockA} \ee
Here $k=-i\epsilon/\hbar-0.5(\Gamma_{\uparrow}+\Gamma_{\downarrow})$
and $\cdots^*$ designates complex conjugation.
$\Gamma_{\uparrow}$ and $\Gamma_{\downarrow}$ designate bath-assisted
uphill and downhill transfer rates in our two-level system calculated as
if, formally (owing to the above scaling) $I=J=K=0$. Let us mention that
via $\Gamma_{\uparrow}$ and $\Gamma_{\downarrow}$, the (initial) bath
temperature $T$ enters the game. These rates are known from the usual Pauli
Master Equation approach to general kinetic problems and, consequently,
are related to each other by the detailed balance condition
$\Gamma_{\uparrow}=\Gamma_{\downarrow}{\rm e}^{-\beta\epsilon}$,
$\beta=1/(k_BT)$. In
general, however, owing to intermixture of coherent and incoherent transfer
channels, the detailed balance condition does not apply to our particle
transfer problem. Finally, $2\Gamma$ is the dephasing rate determined by
$H_{S-B}''$. As for the block ${\cal B}$,
it is fully diagonal with diagonal elements ${\cal B}_{11},\ldots, {\cal
B}_{99}$ equal to $\Gamma_{\uparrow}$, $\Gamma_{\downarrow}$,
$\Gamma_{\uparrow}$, $(\Gamma_{\uparrow}+\Gamma_{\downarrow})/2$,
$\Gamma_{\uparrow}$, $(\Gamma_{\uparrow}+\Gamma_{\downarrow})/2$,
$(\Gamma_{\uparrow}+\Gamma_{\downarrow})/2$, $\Gamma_{\uparrow}$ and
$(\Gamma_{\uparrow}+\Gamma_{\downarrow})/2$, respectively. Next,
\[{\cal D}=\] \[\left(\begin{array}{ccccccccc}
-\Gamma_{\uparrow} & 0 & 0 & iJ/\hbar &iK/\hbar& -iJ/\hbar & 0
&-iK/\hbar&0\\
0 & -\Gamma_{\downarrow} & 0 & -iJ/\hbar & 0 & iJ/\hbar & 0 & 0 & 0\\
0 & 0 & -\Gamma_{\uparrow} & 0 &-iK/\hbar& 0 & 0 &iK/\hbar& 0 \\
iJ/\hbar & -iJ/\hbar & 0 & k^*-2\Gamma & 0 & 0 & 0 & 0 &-iK/\hbar\\
iK/\hbar& 0 &-iK/\hbar& 0 & -\Gamma_{\uparrow}-2\Gamma& 0 & -iJ/\hbar
& 0 & 0 \\
-iJ/\hbar & iJ/\hbar & 0 & 0 & 0 & k-2\Gamma &iK/\hbar& 0 & 0 \\
0 & 0 & 0 & 0 & -iJ/\hbar &iK/\hbar& k-2\Gamma & 0 & 0 \\
-iK/\hbar& 0 &iK/\hbar& 0 & 0 & 0 & 0 & -\Gamma_{\uparrow}-2\Gamma
& iJ/\hbar \\
0 & 0 & 0 &-iK/\hbar& 0 & 0 & 0 & iJ/\hbar & k^*-2\Gamma
\end{array}\right).\] \be \label{BlockD} \ee
As for the block ${\cal C}$, it reads as ${\cal B}$ except for the
interchange $\Gamma_{\uparrow}\leftrightarrow \Gamma_{\downarrow}$.
The form of all the matrices in (\ref{Equat2})
is {\em exactly} the same as, e.g., that one which we would get from the
stochastic Liouville equation SLE \cite{Rein} provided, however, that
$H_{S-B}$ is replaced by a
proper stochastic (e.g. Gaussian delta-correlated) potential field
acting on the central system. The only difference between our form of the
${\cal A}-{\cal D}$ blocks and that of the same matrices in the corresponding
SLE theory is dictated by physics of the problem: Namely, in contrast to
the SLE approach, spontaneous processes with respect to the bath naturally
appear in our fully quantum model ($H_{S-B}$ is, in our case, a coupling to
a genuine {\em quantum} bath). Thus, $\Gamma_{\uparrow}<$ or even
$\ll\Gamma_{\downarrow}$ in our case. As SLE is quite standard and
well understood, this comment will hopefully turn attention of suspicious
readers from potential speculations about correctness of our approach to the
form of our Hamiltonian. It is the latter what is responsible for the
striking conclusions obtained.

{\bf 4. Stationary flow and power output.}

Upper half of (\ref{Equat2}) provides a set of 18 linear algebraic
homogeneous
equations for 18 components of the density matrix $\rho_{uu}$ and $\rho_{dd}$
(remaining components becoming decoupled and irrelevant in what follows) of
the rank 17. So, it should be complemented by the normalization
condition
\be \sum_{m=-1}^{+1}[\rho_{mu,mu}+\rho_{md,md}]=1. \label{norc} \ee
Then this set and (\ref{norc}) determine the relevant components of
the density matrix uniquely. From the physical meaning
of the transfer rates, one can then determine the flow
`-1'$\rightarrow$`0'$\rightarrow$`+1'$\rightarrow$`-1' as
\be {\cal J}=\Gamma_{\downarrow}\rho_{0d,0d}-\Gamma_{\uparrow}\rho_{0u,0u}.
\label{flow} \ee
Calculated flow is, as simple inspection or numerical solution shows,
always nonzero. Already this implies, owing to the persistent character of
the flow, important consequences. We are, however, here interested
rather in the possibility of converting the heat from the bath into a
usable work. For that, we connect the particle running through our
system organized as a circle of three sites (as
`-1$\rightarrow$0$\rightarrow$+1$\rightarrow$-1') with a screw (with its
axis perpendicular to the plane of our three sites `-1', `0' and `+1').
The above persistent flow could then drive the screw at the cost of just
the thermal energy of the bath, converting thus the latter directly to the
mechanical work. So the existence or nonexistence of the stationary flow
according to the above definitions converts to a question of a direct
violation of the 2nd law of thermodynamics as applied to our
system in the sense of the existence or nonexistence of the perpetuum mobile
of the second kind. The problem can be easily solved numerically in our
model. The idea is to provide the system with potential steps the particle
(and the screw) must overcome when passing from `-1' to `0',
from `0' to `+1' or from `+1' to `-1'. (When passing virtually in the
opposite directions, the potential steps as felt by the particle would get
the opposite sign.)
We have chosen these steps equal, designating their value as $\Delta/3$.
Hence $\Delta>0$ is the mechanical work
the particle exerts on the screw during one turn in the above direction. One
should realize that taking these steps as those of the potential energy of
the screw connected with the particle, the latter potential energy is not
unique as a function of the particle position on the above triangle of the
sites. Instead, it is unique as a function of the rotation angle $\phi$ of
the particle (or screw). In other words, when the particle performs one turn
`-1'$\rightarrow$`0'$\rightarrow$`+1'$\rightarrow$`-1', $\phi\rightarrow
\phi+2\pi$, the potential ascribed to the particle connected with the
screw increases by $\Delta$ though the particle formally returns to the same
site.

\begin{figure}
\hspace{1cm} \includegraphics[width=120mm] {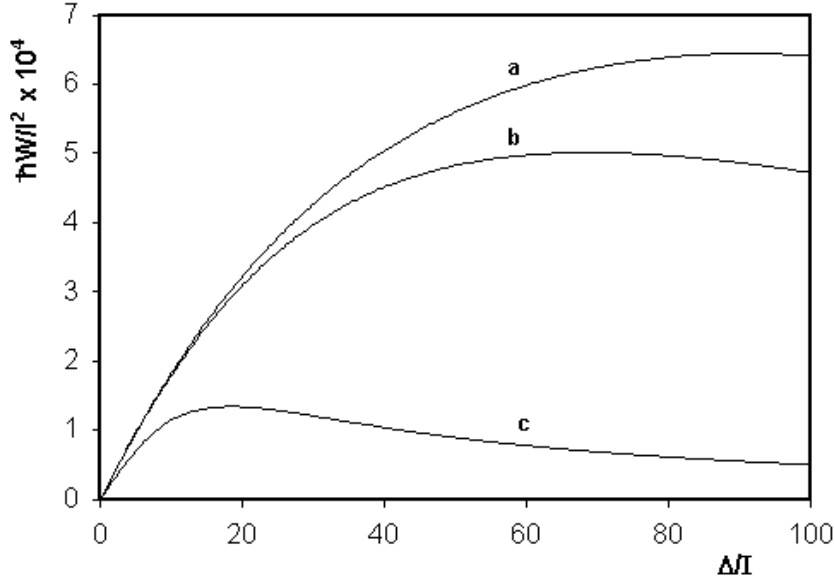}
\caption{Thermal-to-mechanical energy conversion power-output $W$ in
units $I^2/\hbar$ (as determined by (\ref{output}))
as a function of $\Delta$; $I=J>0$, $\epsilon=50I$, $\Gamma_{\uparrow}=0$,
$\Gamma_{\downarrow}=10I/\hbar$, $2\Gamma=3I/\hbar$.
Parameter $K=0.1I$, $0.05I$ and $0.01I$ for curves a, b, and c, respectively.}
\end{figure}

Technically, inclusion of the above potential steps is simple. As it
follows from the above formalism or already even from the Liouville equation
for the whole complex of the system and the bath, it only means to add to
block matrices ${\cal A}$ and ${\cal D}$ above the $9\times9$ block
${\cal E}$ with all matrix
elements equal to zero except for ${\cal E}_{44}={\cal E}_{77}=
{\cal E}_{88}=-{\cal E}_{55}=-{\cal E}_{66}=-{\cal E}_{99}=i\Delta/(3\hbar)$.

From the solution to
\[0=\left(\begin{array}{cc}{
\cal A}+{\cal E} & {\cal B} \\ {\cal C} & {\cal D}+{\cal
E}\end{array}\right)\cdot
\left(\begin{array}{c}\rho_{uu} \\ \rho_{dd} \end{array}\right).\]
\be \label{Equat4} \ee
and (\ref{norc}), one can then determine the mechanical energy power output
(as measured on the screw), i.e. the thermal-to-mechanical energy conversion
power output (per second)
\be W={\cal J}\cdot\Delta. \label{output} \ee

Fig.1 then unambiguously illustrates the positive result of our test
whether the conversion of the thermal energy of our single bath to the
mechanical work is, in our model, possible or not. One should realize the
fact that the mechanical output is, as Fig. 1 shows, really positive.
Hence, the system produces, owing to its selforganizational properties and
properly timed opening and closing of the particle transfer channel across
site `0',
positive mechanical work. This work can, because of the construction of the
model, be at the cost of just heat of the thermodynamic bath. Hence, we
have a direct `heat $\rightarrow$ mechanical work' conversion. As we have
just one bath, there is no compensation possible. Thus, our system working
cyclically provides an example of perpetuum mobile of the second kind,
violating thus the second law of thermodynamics in its Thomson form.

\end{document}